# LUMINESCENCE OF LOCALIZED STATES IN OXIDIZED AND FLUORINATED SILICA GLASS

A.N. Trukhin[1]

Institute of Solid State Physics, University of Latvia, Latvia

Abstract

The photoluminescence of oxidized and fluorinated nominally pure silica glasses, which have good optical transparency in the optical gap, has been investigated under excitation by excimer lasers. The data are compared with the data for chlorinated silica glass, as well as glass containing OH. The excitation of localized states provides the release and capture of charge carriers, despite the low absorption coefficient. Holes are self-trapped. The recombination in the thermally activated or tunnel process creates luminescence of oxygen-deficient centers. It is concluded that oxidation and fluorination do not change the concentration of localized states, but they change the electronic transitions of localized states to higher energies. In type III glass with a high OH content, luminescence is suppressed by the presence of hydrogen and appears after irradiation with photons of 7.9 eV of the $F_2$ laser. This can also be the evidence of existing of localized states even in such glasses.

1. Introduction

Localized states are a specific property of an disordered system. Paraphrasing the well-known definition of localized states made by Mott [1], we can say that in the case of optical glasses, localized states are defects (traps) of the basic structure of glass. Such a definition should contain additional comments. The value of the category "defect" should be wider than a point defect. Mott stated that there is no transport between these traps (defects), despite overlapping wave functions. The overlap of wave functions means that the concentration of defects due to disorder is high.

---

[1] Corresponding author.
E-mail address: truhins@cfi.lu.lv

Currently, the problem lies in the fact that localized states and intrinsic point defects are still not well distinguished, and there are several models for them. Thus, in the energy band model, localized states are described as traps located below the conduction band and above the valence band, having an increased density of states near the bottom of the conduction band and the top of the valence band. A point defect in this approach is a single level in the band gap.

This model works quite well for semiconductor materials [1] and, although it is applicable to insulating materials, however, it is not sufficient to describe the optical properties. For disordered insulating materials, it is necessary to specify the structural features in order to describe the optical properties. Thus, for soda-silicate glasses, localized states belong to the glass modifier sublattice. This concept is crucial for describing the complex properties of luminescence and the effect of radiation [2]. In contrast to sodium germanate glasses, localized states belong to the glass-forming network. This distinguishes the luminescent and radiation properties of sodium-germanate glasses from those of silicate glasses.

Moreover, in the case of sodium germanate glasses, the intrinsic absorption threshold is very close to the threshold of a rutile-germanium dioxide crystal. Even in the case of pure $GeO_2$ glass, there are two absorption thresholds. One at low energy (4.7 eV) coincides with alkali-germanate glasses, and the other at 6 eV almost coincides with the threshold of a germanium dioxide α-quartz crystal (see, for example, [2]). Based on these data, it was concluded that the localized states of germanate glasses are structurally associated with patterns with rutile or octahedral structures. This runs counter to the idea that the octahedral structure is forbidden in disordered materials. We can propose another model of a localized state in terms of the atomic structure. That is, taking into account that an disordered structure can be represented as density fluctuations, a part of the patterns can be the motifs of an octahedral structure having an interface between the main structures (tetrahedral in the case of the glasses under consideration).

Another aspect of localized states is related to the conditions of glass sintering. In the case of soda-silicate glasses, the effect of melting conditions influence on the properties of glass was observed. Reducing melting conditions cause the creation of additional optical absorption below the characteristic oxidized glass optical threshold. For this additional absorption, the excited luminescence is similar to the luminescence excited at a higher energy in the oxidized sample. This approach has been used to explain the properties of silica glass.

The situation of changes in the properties of fused silica in redox conditions has been studied quite well; see, for example, [3, 4, 5]. The main feature of this study is the concept of centers with oxygen deficiency in fused silica glass obtained under reducing conditions. The properties of oxygen-deficient silica glass can be summarized as follows. First, a two-coordinated silicon luminescent center (oxygen deficient center (ODC (II)) was distinguished [6]. This is an example of a point defect in a glass lattice whose properties are usually observed for crystals — exponential decay of luminescence, intra-center thermal quenching with monoenergetic activation. The absorption band at 7.6 eV, often observed in silicon dioxide, was also associated with oxygen deficiency (ODC (I)). Although many scientists consider this band as a point defect - an oxygen vacancy, the detailed studies of the properties show that this band cannot be a point defect [7]. It is rather a complex of defects, [3]. The model of localized state for silicon dioxide, should be similar to the model of localized state in $GeO_2$ glass [2, 8]. It should be based on the fact that the octahedral modification of silicon dioxide crystal, stishovite, has a higher intrinsic absorption threshold than α-quartz, [9]. The corresponding stishovite-like structural motifs are located in the region of silica host absorption of a basic structure and cannot be directly detected optically. However, in the case of oxygen deficient samples, localized state can be observed by analogy with the case of silicate sodium and phosphate glasses deficient in oxygen [2]. The stishovite luminescence is very similar to the luminescence of oxygen-deficient quartz glass. Also, the absorption band at 7.6 eV was observed in stishovite single crystals [9]. The oxygen deficiency of silica ultimately stimulates the manifestation of a localized state at energy below the intrinsic absorption threshold.

Earlier, localized states were studied in silica samples with a strong oxygen deficiency and, accordingly, high optical absorption in the optical gap [10]. The main conclusion of these studies is the excitation of recombination luminescence of an electron trapped at ODC and a self-trapped hole, [11, 12]. These traps are localized states for electrons below the conduction band and localized states for holes above the valence band.

The problem lies in the properties of localized states in pure silica obtained under normal or oxidized conditions. There may be traces of states located below the optical gap, and the search and study of these states is the task of this work. Such low density states can play a crucial role in the elementary processes responsible for initiating many photostimulated reactions. For example, solarization of optical fibers produced from silicon dioxide.

Glasses of pure silica type IV, manufactured by KC4B technology [13], were sintered in conditions of excess oxygen or fluorinated. These samples have an extremely low absorption band at 7.6 eV. The comparison was carried out with a sample containing chlorine, also manufactured by the KC-4B (or KS-4V) technology, with a relatively high intensity absorption band at 7.6 eV. Samples of type III silica containing a large amount of OH groups and type IV samples activated by aluminum, but without alkali ions, were also investigated. The latter sample was chosen because Al at ambient temperature effectively captures the holes [14], whereas at low temperature the holes are mostly self-trapped [14]. It must be emphasized that the self-trapped hole is a unique property of quartz glass. There is no analogous self-trapped hole in a pure α-quartz crystal. The discovery of self-trapping of holes in silica by the method of electron spin resonance (ESR) [14] is a very significant achievement for the physics of localized states of silica. It directly indicates a state located near the upper part of the valence band [14]. The corresponding captured electron has not yet been detected by the EPR method, but studies of luminescence recombination suggest that electron capture is associated with ODC [15].

2. Experimental procedure

Samples of nominally pure quartz glass with a low content of hydroxyl were investigated by luminescent and photoelectric methods. These samples were produced using the method of KS-4V, developed in [13]. The metal impurity level is about $10^{-6}$ % by weight, and the content of OH groups is about $10^{-7}$ % by weight, [13]. The preparation process was based on the electric melting of cristobalite. For the purpose of purification the crystallization of silicon dioxide was carried out in the presence of a catalyst (lithium), which was then washed out with chlorine. Chlorine was then removed with oxygen. Some samples were obtained without an oxygen washing step, then the sample contained 0.1 % chlorine [13]. Doping of F was performed by melting in an atmosphere of pure $SiF_4$ [16]. The concentration of fluorine in the doped sample is 0.1 wt. %. Samples melted in an oxygen atmosphere are considered oxidized. The content of OH groups in pure quartz glass type III (Corning 7940 and KY-1) was about $10^{-1}$ % by weight. Samples were polished and annealed at 500 °C before measurements. For studying the low-density states in silica, UV excimer lasers were chosen as excitation tools. A krypton-fluorine laser has an emission wavelength at 248 nm, a fluorine-argon laser - at 193 nm and a fluorine laser - at 157 nm. These photons correspond to the energy range below the optical gap of quartz.

The photoluminescence spectra and the decay kinetics curves were measured in the temperature range from 80 to 400 K.

3. Results

Photoluminescence of type IV water-free silica

The optical absorption spectra of the samples under study are shown in Fig.1. It has been established that the main samples under investigation, fluorinated and oxidized, have the lowest degree of optical absorption, which is close to the intrinsic absorption threshold at 8.1 eV. The sample containing chlorine has a well-defined band at 7.6 eV. A sample of type III silica with a high OH concentration does not have a band expressed at 7.6 eV; however, in this sample we observe a strong absorption, which significantly shifts the absorption threshold to 7.8 eV.

Excitation with a 157 nm laser ($F_2$) provides luminescence from 220 to 700 nm with a main band at 450 nm and a shoulder in the UV range, Fig.2. The decay kinetics is long for the blue and ultraviolet spectral ranges, Fig.2, the right insert for the blue and the left insert for the ultraviolet part of the spectrum.

Fig. 3 shows the time-resolved spectra of photoluminescence (PL) and time constant in oxidized silica excited at 193 nm and 157 nm. Intensities of PL are obtained by integration of decay curve measured at corresponding wavelength. The stationary spectrum of the same sample of silicon dioxide excited at 160 nm by a deuterium light source through a vacuum monochromator, is also presented for comparison. The latter spectrum is the usual silica ODC luminescence spectrum. Time-constant spectra are obtained by an exponential approximation of the decay curves. The corresponding decay curves are expressed in hundreds of μs for the blue part of the spectrum and in tens of μs for the ultraviolet part. In addition, the UV band has a very fast decay component, Fig. 4, which can be approximated with a time constant of about 1-2 ns. Such behavior was previously also observed for silica samples with a higher oxygen deficiency with well-pronounced ODC (II) bands (~ 5 eV) [4] and ODC (I) (7.6 eV) [3].

The result of thermally stimulated luminescence (TSL) is demonstrated in figure 5. The case of excimer laser excitation is compared with X-ray irradiation of samples at 80 K. It is observed that the TSL curves are wide and are in the temperature range from 100 to 220 K. Two bands of blue and ultraviolet light are presented in TSL. TSL is most efficiently generated with an X-ray and $F_2$ laser (157 nm).

The correspondence between the thermal dependences of the photoluminescence intensity and its time constant, on the one hand, and the thermal stability of the STH signal, on the other, is remarkable [15], Fig. 6. The decrease in the intensity of ESR signal corresponds to the release of STH [15]. It can be assumed that recombination luminescence is associated with recombination between STH and an electron trapped at an ODC-type defect. The release and separation of a hole from trapped electrons leads to a decrease in the PL intensity. In addition, this process causes an acceleration of the recombination rate and a decrease in the luminescence duration shown in Fig.6 by open dots. Already studied TSL, as well as EPR measurements on samples with oxygen deficiency [12] show that the nature of TSL peaks is determined by the release of self-trapped holes (STH ) of two types. Then in oxidized and fluorinated samples, a similar [12] TSL mechanism is observed. The tunneling character of recombination luminescence can be demonstrated using the decay kinetics of $t^{-1}$, observed for cases of excitation by 157 nm and 193 nm of excimer lasers, Fig. 7. The correspondence of such kinetics for tunnel processes was established in [17].

The decay kinetics law is similar for excitation with these lasers, despite the large difference in the energy of the corresponding photons: 7.9 eV for the $F_2$ laser and 6.4 eV for the ArF laser. It differs when luminescence is excited with a KrF laser (248 nm or 5 eV). A comparison of the PL spectra and the decay kinetics is shown in Fig. 8. The luminescence duration, when excited by a KrF laser, is much longer than when excited by an ArF laser. Also, in the case of excitation by a KrF laser, the intensity of the blue PL decreases differently than when excited by 193 nm and 157 nm of lasers, Fig. 9. Differences are observed in the PL spectra shown in Fig. 8. Visual observation of an excited PL using a KrF laser shows that the blue luminescence is relatively bright near the sample surface. In the volume, the red luminescence of the non-bridging oxygen luminescent center is observed mainly for the oxidized sample. However, in the measured spectra, the red luminescence is less intense than blue, Fig.10. The red band is practically not observed for fluorinated and chlorinated samples under the same excitation conditions. The UV-PL band is also observed in oxidized and chlorinated samples, whereas for fluorinated samples the UV-band is poorly resolved.

The intensity of blue luminescence decreases exponentially with increasing temperature, whereas the shape of the decay kinetics does not depend on temperature, Fig. 11, when excited with a KrF (248 nm) laser. In the case of $F_2$ and ArF lasers, the luminescence duration decreases with increasing temperature. Thus, the nature of luminescence is at least partially different for KrF excitation, on the one hand, and $F_2$ and ArF lasers, on the other. We assume that a KrF laser can excite the luminescence of a self-trapped exciton (STE) in silica glass, as was obtained earlier [18].

Photoluminescence of type III silica containing OH

As is known, the presence of OH groups in silica modifies ODC killing their ability to provide luminescence [19]. However, continuous irradiation resumes luminescence ability [3]. A similar effect is observed in the present experiment. Fig.12 shows an example of such a recovery of the luminescence. A sample of type III silica (Corning 7940) is continuously irradiated with $F_2$ laser pulses (157 nm), and the luminescence spectra are measured changing the monochromator wavelength point by point. Registration of one point was 1 second; thus, for the first measurement of the spectrum from 200 nm to 800 nm, the sample was irradiated for 135 seconds. The next spectrum began to be measured almost immediately after the first one had already noticeably increased the intensity. The luminescence spectrum is wide from 220 to 800 nm with a well-pronounced band of about 650 nm. The measured decay kinetics curves of the 650 nm band show that the red band is the luminescence of non-bridging oxygen center created by laser irradiation, Fig.13. Detailed information on the center of the luminescence of non-bridging oxygen is available in [20]. This center cannot be created by an ArF laser (193 nm) in samples annealed at 500 ° C. However, after irradiation with $F_2$ laser, red luminescence is also observed when ArF is excited. The blue center of luminescence is created by both lasers in silica type III. However, luminescence in the ultraviolet part of the spectra is more efficiently excited by the $F_2$ laser, but it is also observed when ArF is used, Fig.14 (kinetics for 300 nm). At low temperature (80 K), the situation is similar to 293 K, Fig.15 (a small band at 300 nm for excitation with the ArF laser). Induced luminescence has components of the fast and slow decay, Fig.16. The faster component is more intense in the ultraviolet. As before [3], for water-free and oxygen-deficient sample treated with hydrogen, laser-induced luminescence is assigned to ODC modified with OH groups. A special case of luminescence excitation is observed when a KrF excimer laser (248 nm) is used. At temperature of 293 K, the red luminescence of the center of non-bridging oxygen was observed, Fig. 17. Other luminescence in the spectrum is negligible. The luminescence intensity increases almost exponentially with cooling, fig. 17. Red luminescence at low T is significantly less intense. The decay kinetics curves of the red luminescence excited by a KrF laser are not exponential at 290 K, and are longer than those observed with $F_2$ and ArF lasers, Fig.18. The decay curves of the blue luminescence are in the time scale of ms, and the decay kinetics is almost independent of temperature, despite the strong dependence of the temperature intensity, Fig.18. This case is very similar to the one shown in Fig. 11 for type IV silica. A similar interpretation can be offered. Under the irradiation with 248 nm of laser, the luminescence of the STE is excited. [18].

4. Discussion

Silica samples having a relatively low optical absorption level in the transparency range (below the optical gap at about 8 eV) were investigated by the photoluminescence method. Earlier, the same samples were studied by photoelectric methods [21], where charge release from excited localized states was detected. The amount of optical absorption was reduced by additional oxidation or fluoridation. The release of electric charge and recombination luminescence were observed when samples were excited by KrF (248 nm or 5 eV); ArF (193 nm or 6.42 eV) and $F_2$ (157 nm or 7.9 eV) excimer lasers [21]. The photoelectric data were interpreted as the release of electrons and holes, creation of the Dember field, depending on the properties of the traps for charge carriers, and the current limited by the space charge, [21]. Samples containing effective hole traps (Al at 293 K) provide the negative sign of the electron-induced Dember field. Otherwise, in pure samples at high temperatures > 200 K, the holes move and the electrons are captured, then the positive sign of the Dember field is associated with the moving hole. At low temperatures, below the STH release <180 K, the holes self-trapped, and the negative sign of the Dember field shows a higher electron mobility [21].

Spectrally observed luminescence is associated with ODC with the observation of two main blue and ultraviolet bands. As before [3], the decay kinetics of ODC luminescence differs from the decay kinetics of an ODC (II) point defect or twofold coordinated silicon. The decay kinetics of ODC (II) is exponential with a time constant of 10 ms for the blue band and 4.5 ns for the UV band [4]. In Figs. 2, 3, the decay of the UV band is not exponential and lasted for tens of microseconds - much longer than that of ODC (II). The decay of the blue band is not exponential, that is characteristic of disordered materials, and the duration of the luminescence is shorter than 10 ms. Deviations from exponential decay are explained as recombination excitation of blue luminescence. The decay prolongation for the UV band also is explained by the recombination process. On the other hand, the observed another component of decay is faster than the characteristic 4.5 ns ODC (II) decay for the UV band, Fig.4. It is short, about 1-2 ns, i.e. shorter than for a single ODC (II). The shortening of the decay of the UV band, shown in Fig. 4 may occur due to the structural deviation of the ODC associated with localized states from the ideal twofold coordinated silicon.

The recombination mechanism can be associated with the tunnel recombination between electrons and self-trapped holes. The law of decay kinetics ~ $t^{-1}$, shown in Fig. 7, may indicate tunnel recombination, as it was previously shown [17]. For the deeper traps, recombination with release of charge carriers and their entering into the conduction and valence bands was also observed. TSL after irradiation at low temperatures (80 K) was obtained, Fig.5. It is noteworthy that the curve of TSL is in the range of STH release. It is possible that a thermally released hole moves toward an electron trapped at an ODC located at a distance and recombines with this electron. Thus, the excited localized states provide the release of electrons and holes. The capture of electrons leads to the creation of an ODC with an electron. The hole is self-trapped. The decrease in the intensity of the ESR corresponds to the release of STH [22], Fig.6.

Properties of type III silicon dioxide, containing OH groups, are different from those of type IV silica without OH. For type III silica, luminescent ODC is suppressed by the presence of hydrogen and appears after irradiation with photons of an $F_2$ laser with energy of 7.9 eV, Fig. 12. Notable is the creation of a red luminescent center of non-bridging oxygen during irradiation. Earlier [18], type III silica was used to study the luminescence of self-trapped excitons (STE) under a two-photon excitation of a KrF (248 nm) laser due to the suppression of the ODC emission with OH. The relevant data are shown in Figure 17. It should be noted that the two-photon excitation by a KrF laser practically does not ensure STE emission in the bulk of α-quartz crystal sample, [18]. The STE luminescence was observed on the surface of the crystalline sample even in the single photon mode. In fact, excitation of silica type IV by a photon of the KrF laser in single photon mode provides luminescence with decay kinetics and spectral composition different from the case of excitation by photons of the ArF laser (193 nm), Fig. 8. The thermal dependences of the blue luminescence, excited by 157 nm and 193 nm of excimer lasers, on the one hand, and 248 nm of KrF laser, on the other, Fig.9, differ. That is because KrF mainly excites STE luminescence, whereas $F_2$ and ArF mainly excite ODC luminescence. The decay curves of the blue luminescence of silica type IV excited with 248 nm light, Fig. 11, are practically independent of temperature, but their intensity decreases with increasing temperature. This is indication of STE luminescence observation, [18].

Beside observation of STE luminescence upon excitation using a 248 nm laser light for silica type III and type IV, an ODC emission is observed. This is seen in Figures 16 and 17 as fast UV luminescence for type III silica. The UV band is also observed in the PL spectra of type IV silica excited by a KrF laser, Fig. 10, despite the low absorption coefficient at 248 nm. The blue band of ODC could not be distinguished because it is covered by STE luminescence in condition of 248 nm excitation at 80 K. Temperature dependence of blue PL in type IV silica excited by KrF laser, Fig. 9 does not correspond exactly to type III silica, see Fig. 17, inset. This difference can be explained by the excitation of blue luminescence of ODC by 248 nm laser in type IV silica at 300 K. In type III silica ODC luminescence is not seen when excited with KrF laser. In the same conditions STE luminescence is quenched at 300 K therefore is not observed in type III silica. The luminescence of STE excited in type IV silica using a KrF laser in the one-photon process can be explained by a two-step process involving surface states, as in the case of STE luminescence in α-quartz photons from the KrF laser. Localized states can also provide a two-step process for creating an exciton (~ 10 eV).

5. Conclusions

In the samples studied the ODC luminescence exhibits decay kinetics that differs from those of an ODC (II) point defect or two-coordinated silicon.

Recombination luminescence is associated with recombination between a STH and an electron trapped at an ODC defect. The thermally stimulated luminescence curve is in the range of STH release. The thermally released hole moves and recombines with an electron trapped at ODC.

Therefore, despite the reduced optical absorption of samples with a high content of oxygen or fluorine, the same processes of release, capture and recombination of electrons and holes are observed, as in silica with a high oxygen deficiency. The hole is mostly self-trapped, and electrons are trapped in oxygen-deficient centers.

Acknowledgments
This work was supported by the Latvian Science Council Grant No lzp-2018/1-0289.

References

[1] N.F. Mott, E.A. Davis, Electronic Processes in Non-Crystalline Materials, Oxford (1971) 472 p.

[2] A.N.Trukhin, Excitons, localized states in silicon dioxide and related crystals and glasses, International school of solid state physics, 17th course., NATO science series. II Mathematics, Physics and Chemistry Defects in $SiO_2$ and related dielectrics: science and technolohy, Ed D.Griscom, G.Pacchioni, L.Skuja, Kluwer Academic Publishers, Printed in the Netherlands, 2 (2000) 235-283.

[3] A.N. Trukhin, H.-J. Fitting, Investigation of optical and radiation properties of oxygen deficient silica glasses, J. Non-Cryst. Solids, 248 (1999) 49-64.

[4] L. Skuja, Optically active oxygen-deficiency-related centers in amorphous silicon dioxide, J. Non-Cryst. Solids, 239 (1998) 16-48.

[5] M. Kohketsu, K. Awazu, H. Kawazoe and M. Yamane, Photoluminescence Centers in VAD $SiO_2$ Glasses Sintered under Reducing or Oxidizing Atmospheres, Japanese Journal of Applied Physics, 28 (1989) 615-621.

[6] L.N. Skuja, A.N. Streletsky, A.B. Pakovich, A new intrinsic defect in amorphous $SiO_2$, Solid State Commun. 50 (1984) 1069.

[7] A.N. Trukhin, Radiation processes in oxygen-deficient silica glasses: is ODC(I) a precursor of E'-center? J. Non-Cryst. Solids, 352/28-29 (2006) 3002-3008.

[8] Trukhin A.N., P.A.Kūlis,Lokalized states in germanate glasses. Ultraviolet absorption tail of crystalline and glassy germanium dioxide and alkali germanate, J. Non-Cryst. Solids, 188 (1995) 125-129.

[9] A.N.Truhins, J.L.Jansons, T. I.Dyuzheva , L. M. Lityagina , N. A.Bendeliani, Intrinsic absorption threshold of stishovite and coesite, Solid State Communications, 131 (2004) 1-5.

[10] A.N. Trukhin, Luminescence of localized states in silicon dioxide glass. A short review, J. Non-Cryst. Solids, 357 (2011) 1931–1940.

[11] D.L. Griscom, γ-Ray-induced visible/infrared optical absorption bands in pure and F-doped silica-core fibers: are they due to self-trapped holes? J. Non-Cryst. Solids 349 (2004) 139-147.11

Figures caption

Fig.1

Optical absorption of pure silica glasses type III and type IV. 1 – excess $O_2$, 2 – OH containing type III Corning 7940, 3 - fluorine doped, 4 - $Cl_2$ containing.

Fig.2

PL spectra and decay kinetics of silica type IV samples excited with $F_2$ (157 nm) excimer laser. F- fluorine doped, $O_2$ – oxidized and $Cl_2$- chlorine containing.

Fig.3

PL time resolved and τ spectra of silica type IV (oxidized) excited with ArF and $F_2$ excimer lasers at 77 K. Excitation: 1 – 160 nm of $H_2$ discharge light source, 2 – 157 nm of $F_2$ excimer laser, 3 - 193 nm of ArF excimer laser.

Fig.4

280 nm PL decay kinetics in different samples of silica type IV under pulses of ArF laser.

Fig.5

TSL of silica different KC4B samples after x-ray, $F_2$ and ArF lasers irradiation at 80 K

1-fluorinated sample, x-ray irradiated, blue band at 3 eV; 2- oxygen surplus sample, x-ray irradiated, UV band at 4.4 eV; 3- sample with $Cl_2$, x-ray irradiated, UV band at 4.4 eV; 4- sample with $Cl_2$, irradiated 157 nm light of $F_2$ laser, UV band at 4.4 eV; 5- oxygen surplus sample, irradiated 157 nm light of $F_2$ laser, UV band at 4.4 eV; 6- sample with $Cl_2$, irradiated 193 nm light of ArF laser, 7- oxygen surplus sample, blue band at 3 eV, irradiated 193 nm light of ArF laser.

Fig.6

Correlation of temperature dependences of STH signal [14, 22] and PL intensity & time constant for UV luminescence of KC4B silica type IV under $F_2$ laser.

Fig.7

Peculiarities of blue PL decay kinetics of oxygen surplus sample KC4B silica excited by 193 nm of ArF and 157 nm of $F_2$ lasers at 77 K.

Fig.8

Peculiarities of PL spectra and decay kinetics (insert) of oxygen surplus sample silica type IV excited by 193 nm of ArF and 248 nm of KrF lasers.

Fig.9

Temperature dependence of 430 nm PL oxidized silica type IV excited by excimer lasers.

Fig.10

PL spectra excited by 248 nm of KrF excimer laser in oxidized ($O_2$) fluorinated (F) and chlorine containing ($Cl_2$) silica type IV. Insert - decay kinetics. 80 K.

Fig.11

PL decay kinetics excited by 248 nm of KrF excimer laser at different temperatures in oxidized silica type IV.

Fig.12

Development of PL spectra during $F_2$ laser (157 nm) irradiation in silica type III. 1 –first spectrum, 2-second, 3 –third, 4 –fourth, 5 – fifth spectrum. T=293 K.

Fig.13

PL at 640 nm decay kinetics of non-bridging oxygen center created by $F_2$ laser in silica type III. Insert – development of PL spectra during $F_2$ laser irradiation. ArF data are obtained after $F_2$ irradiation.

Fig.14

decay kinetics PL selected at different wavelengths of silica type III excited by 157 nm light of $F_2$ (a) and 193 nm ArF (b) lasers at 293 K.

Fig.15

Development in time of PL spectra of silica type III excited by 157 nm light of $F_2$ and 193 nm ArF at 80 K.

Fig.16

Time resolved PL spectra of silica type III, excited by 193 nm of ArFlaser, 293 K.

Fig.17

Time resolved (ms and ns) PL of silica III excited by KrF laser (248 nm) pulses at 290 and 80 K. Insert – temperature dependence of time resolved luminescence intensity at 440 nm. Red luminescence at 290 K is luminescence of non-bridging oxygen created by photon stimulated detachment of hydrogen from ≡Si-O-H groups.

Fig.18

PL (at 440 nm) decay kinetics curves at different temperatures of KrF (248 nm) excited silica of III type. Insert: decay kinetics of red luminescence of non-bridging oxygen center.

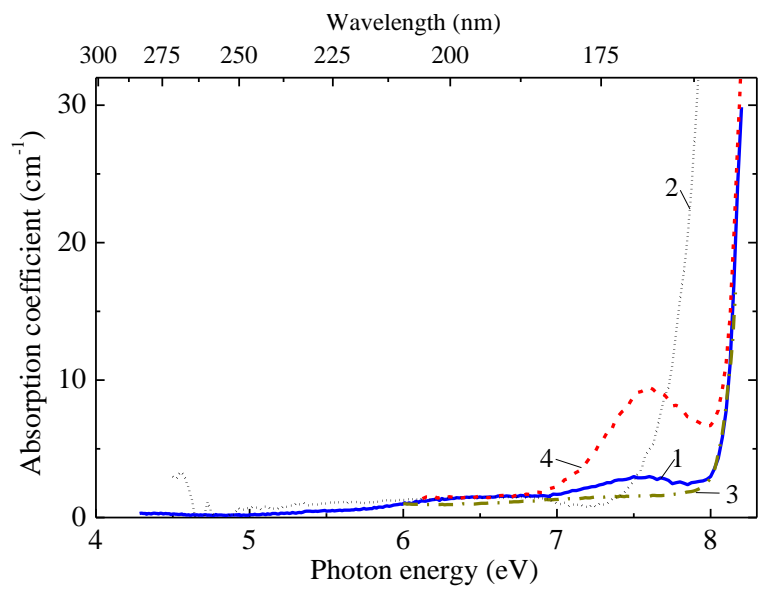

Fig.1

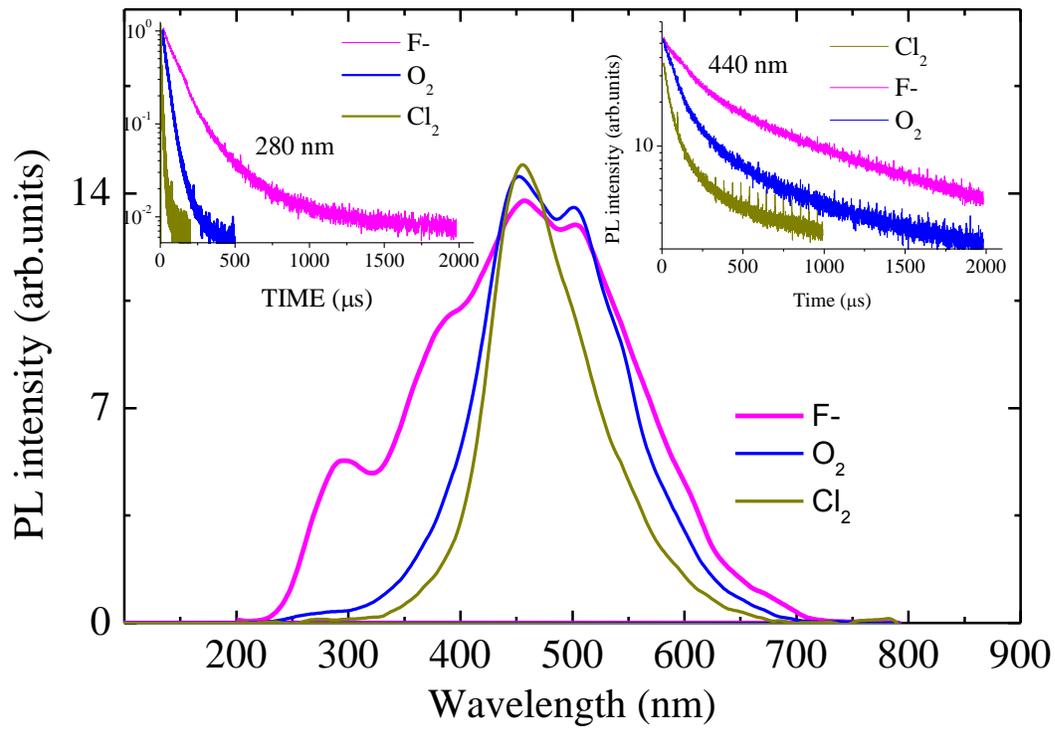

Fig.2

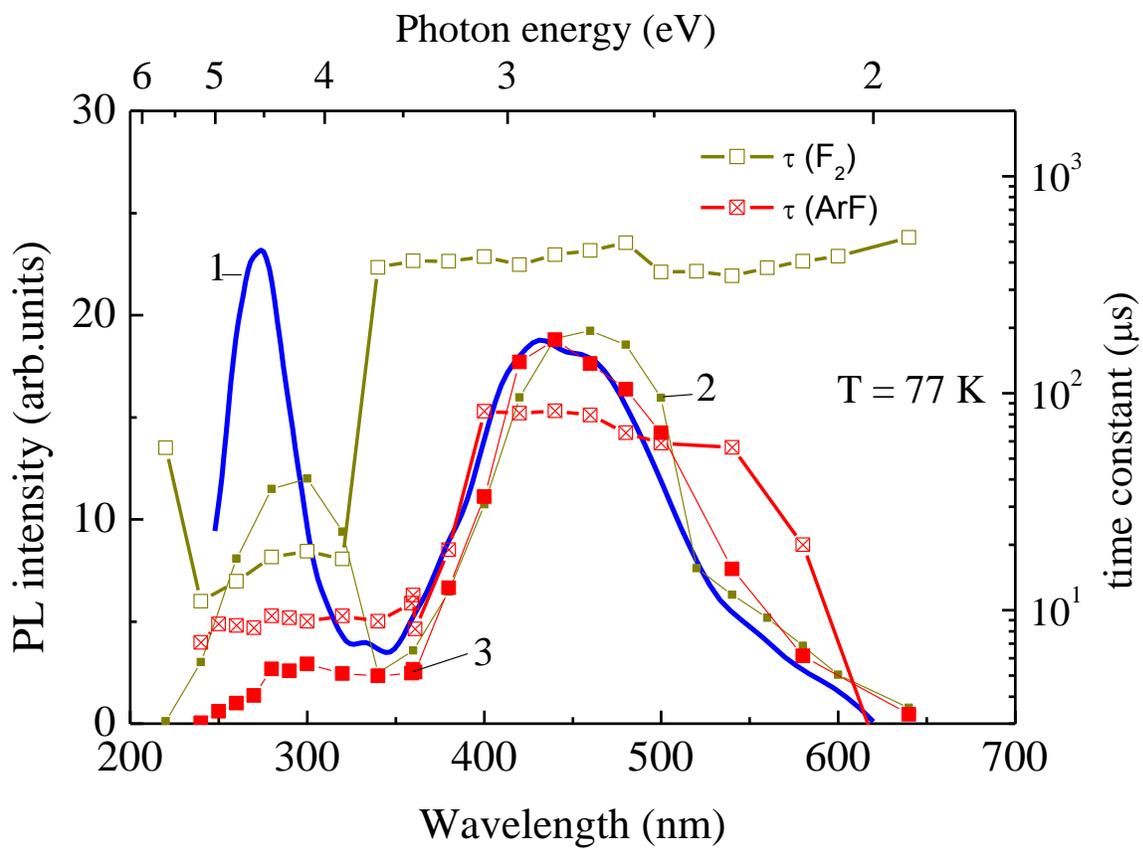

Fig.3

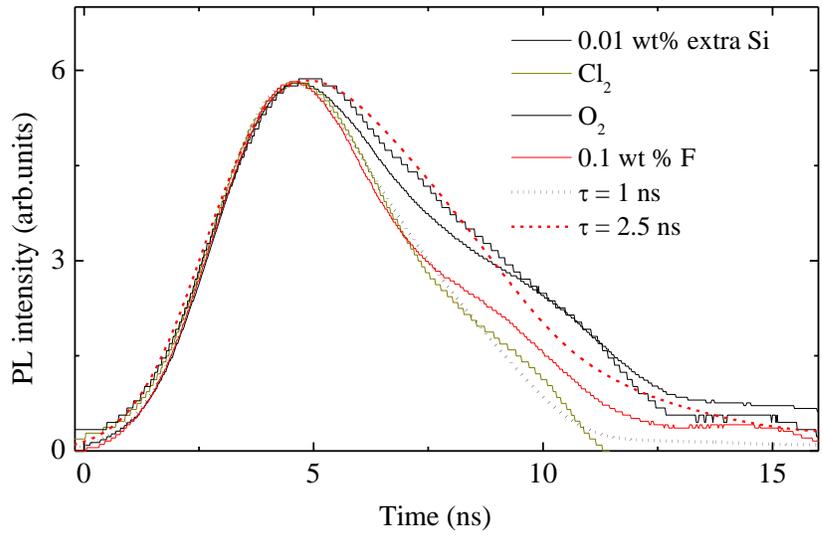

Fig.4

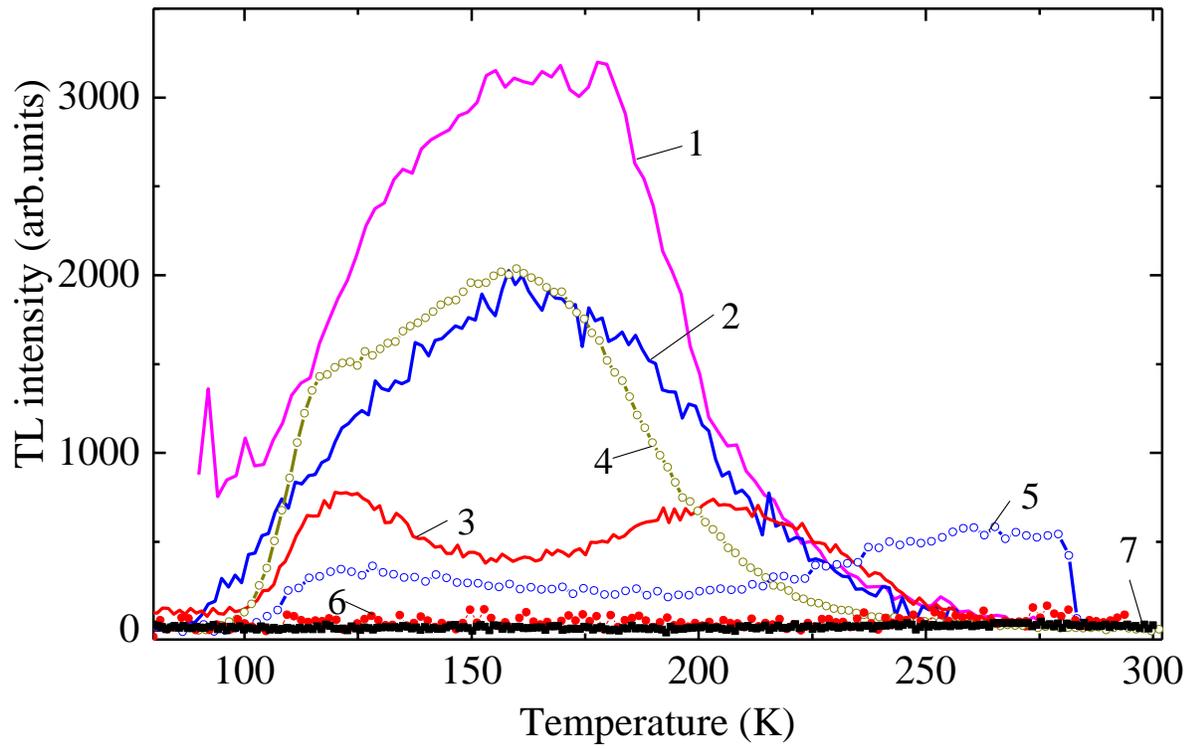

Fig.5

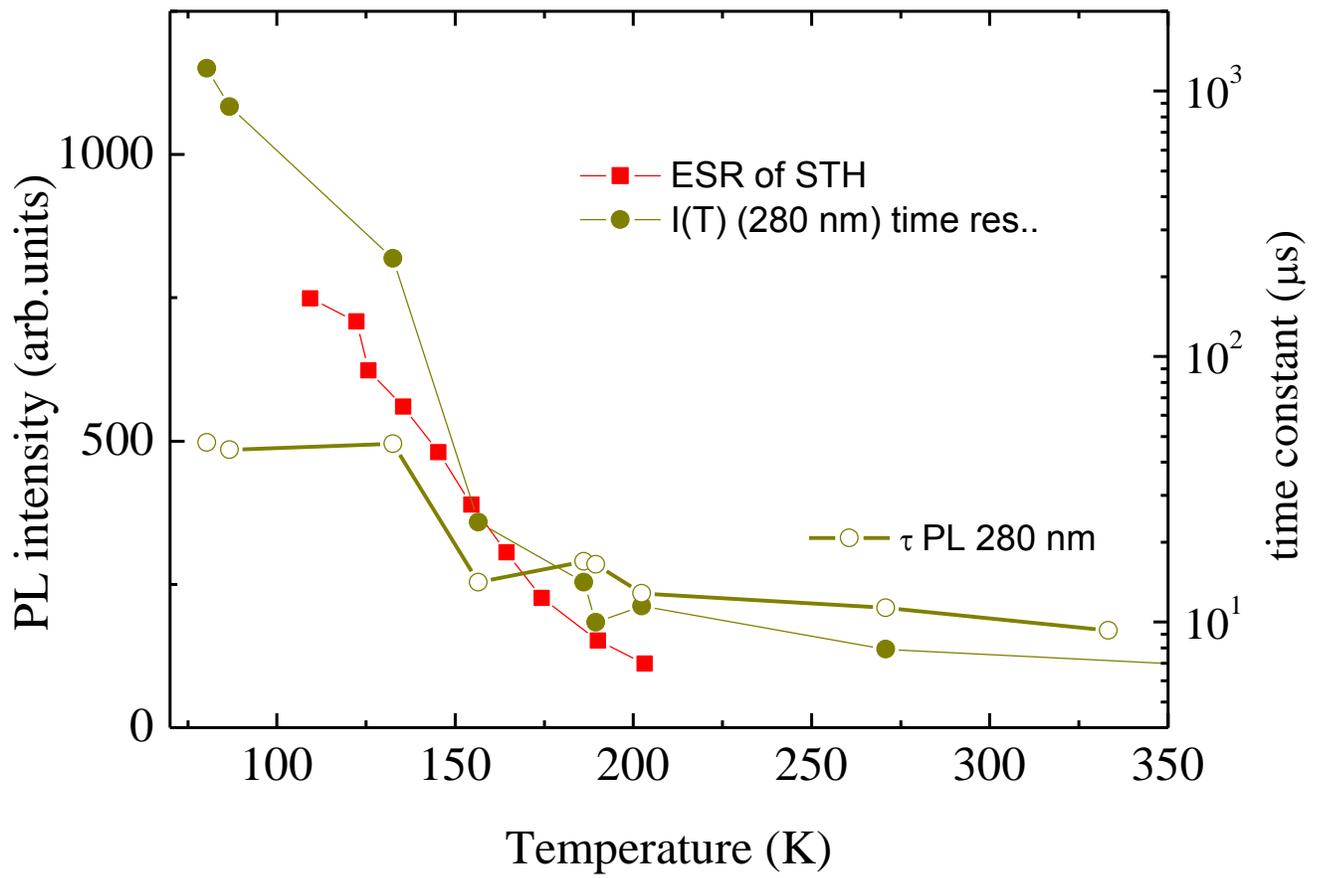

Fig.6

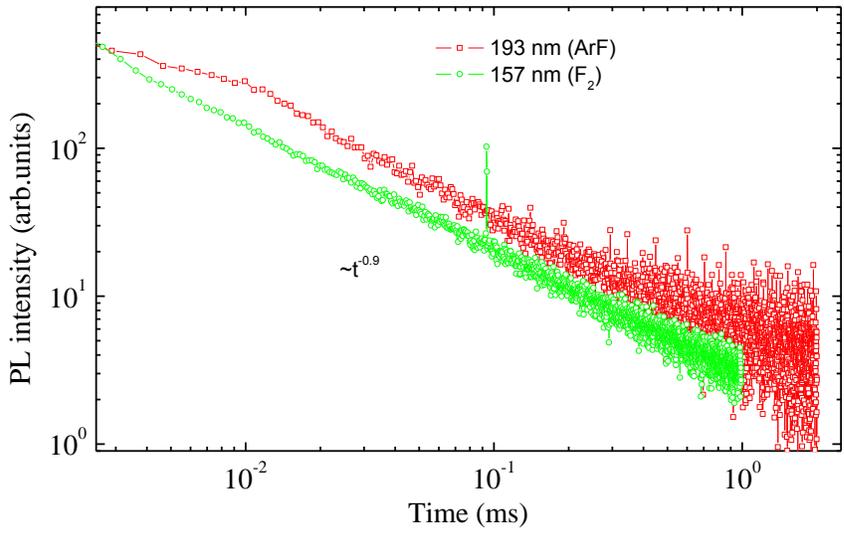

Fig.7

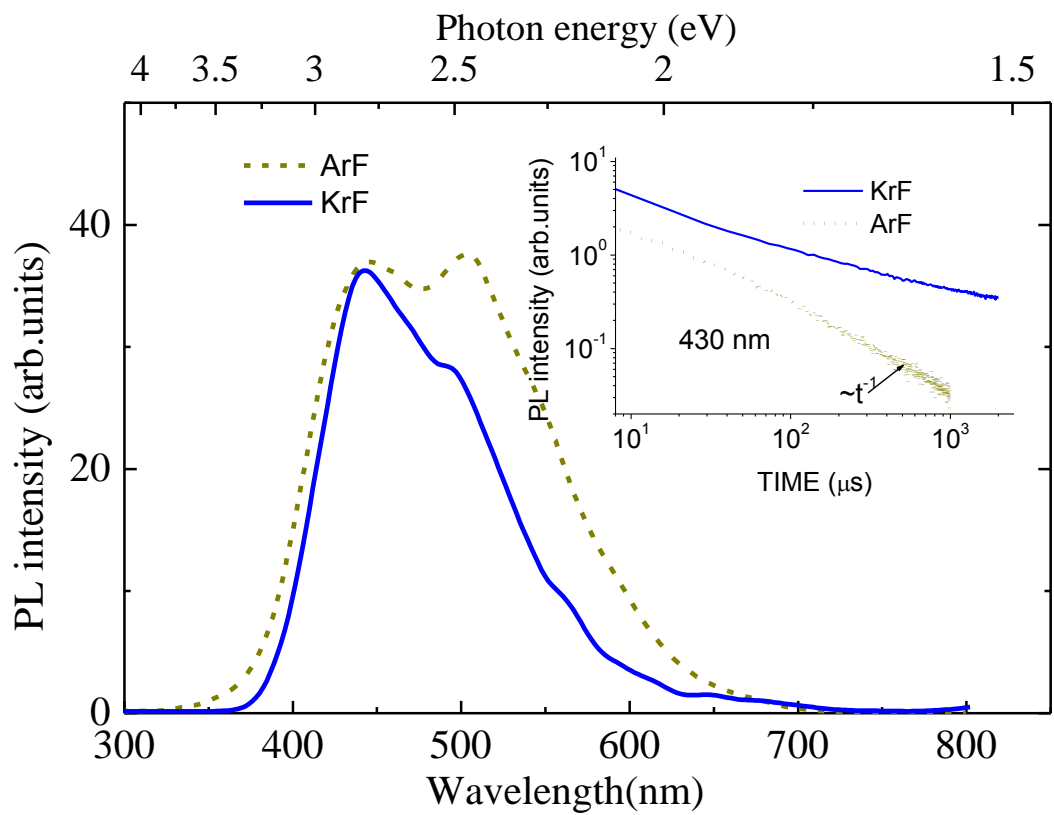

Fig.8

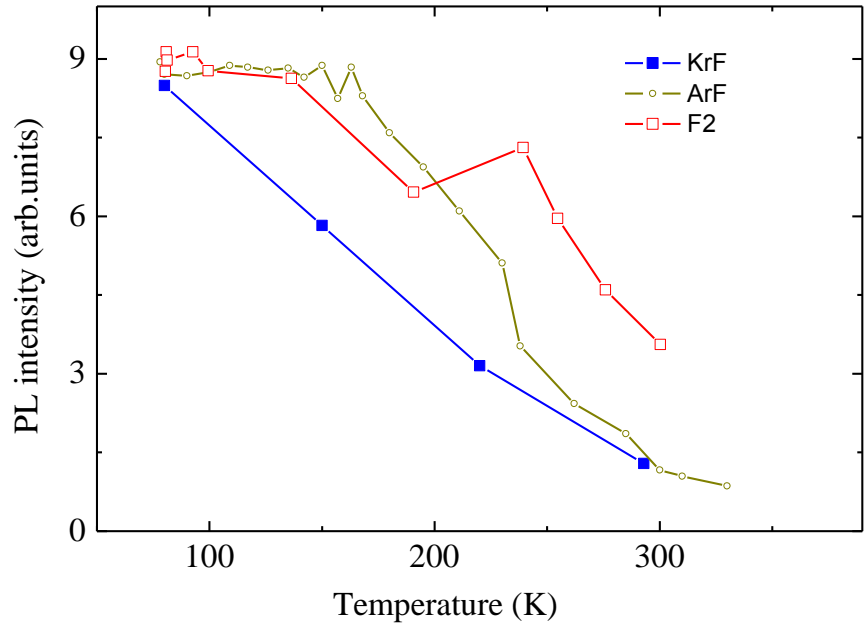

Fig.9

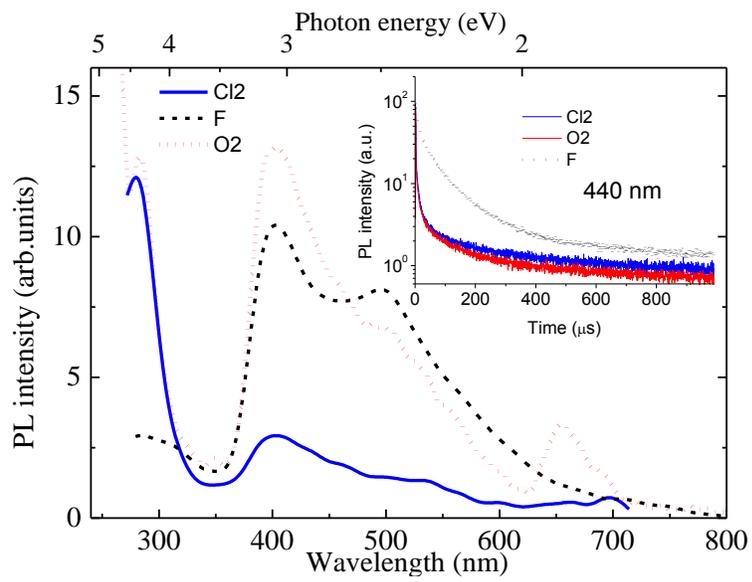

Fig.10

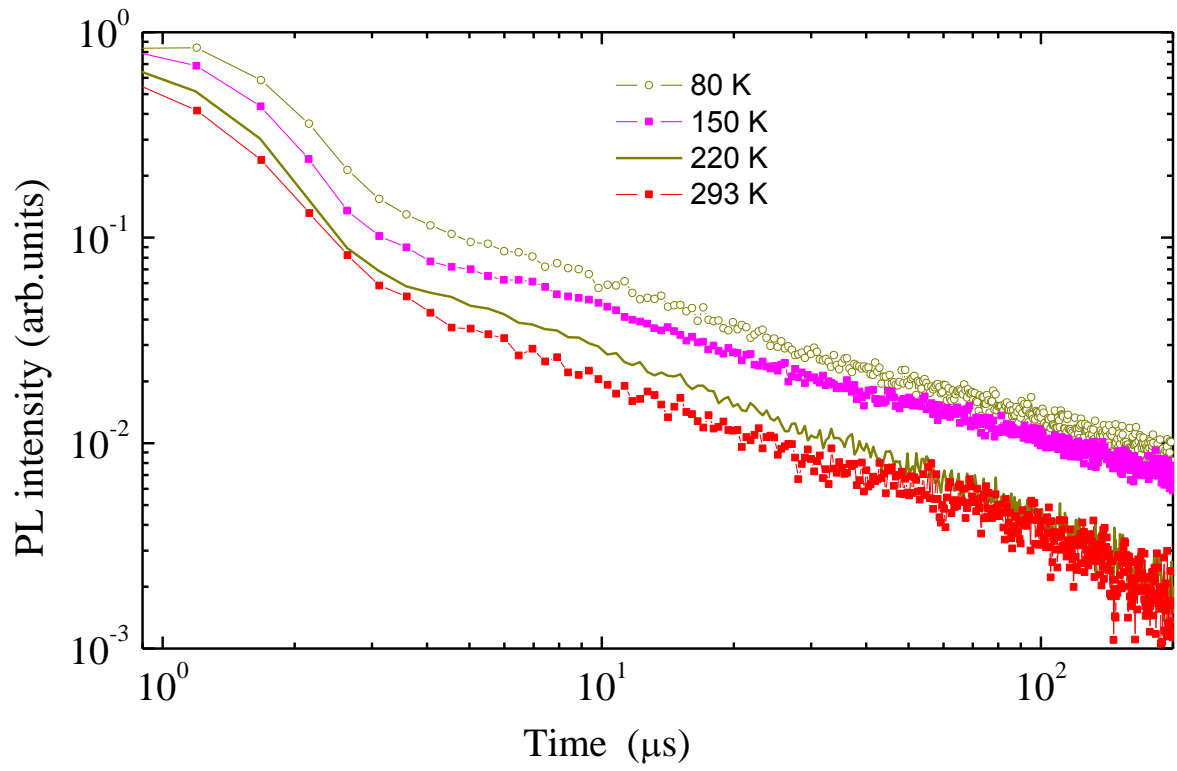

Fig.11

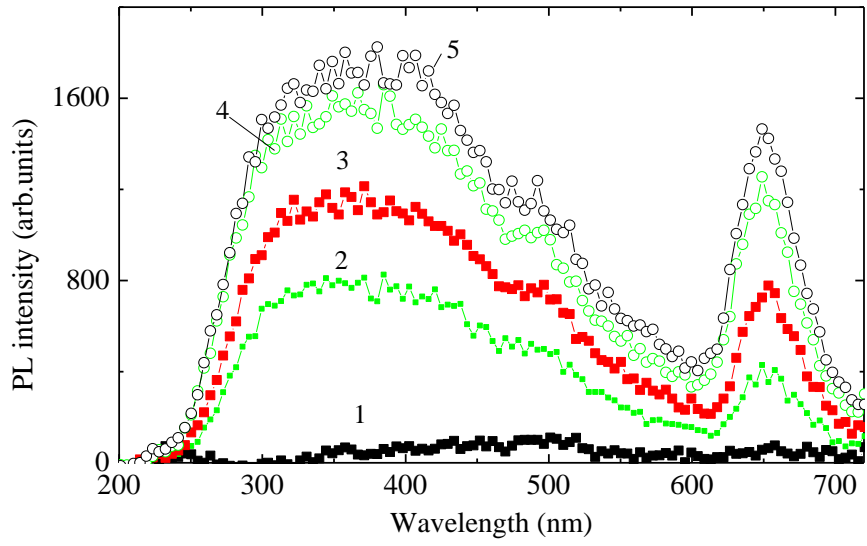

Fig.12

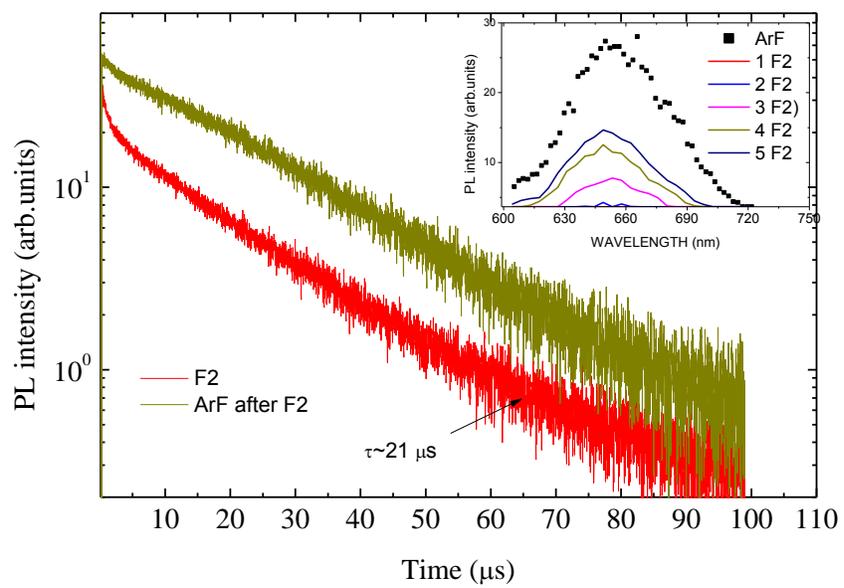

Fig.13

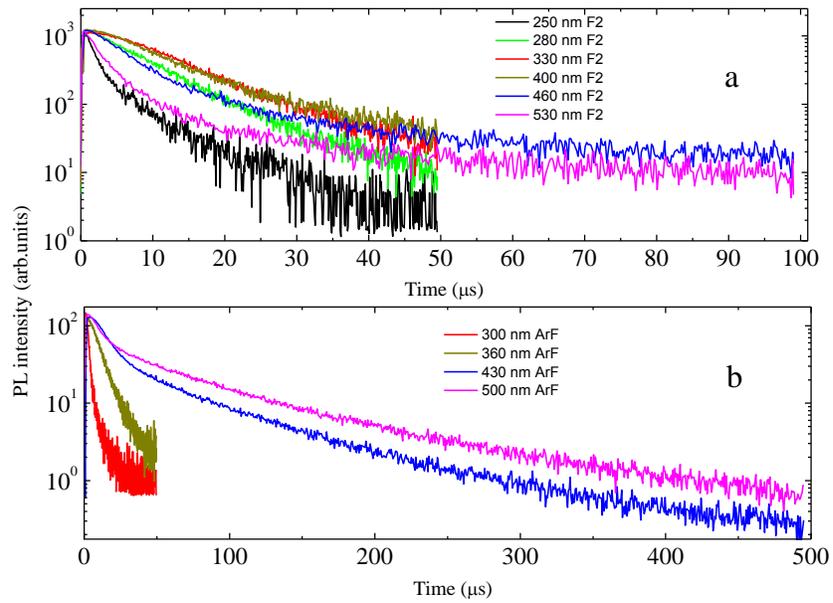

Fig.14

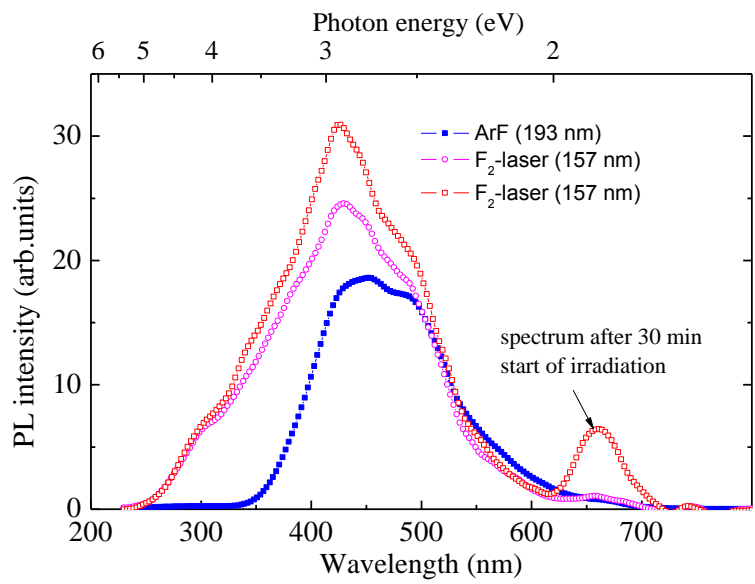

Fig.15

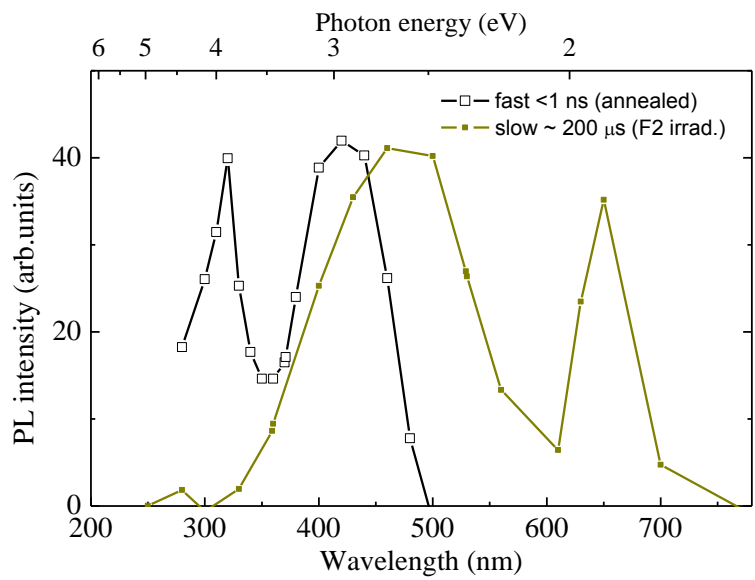

Fig.16

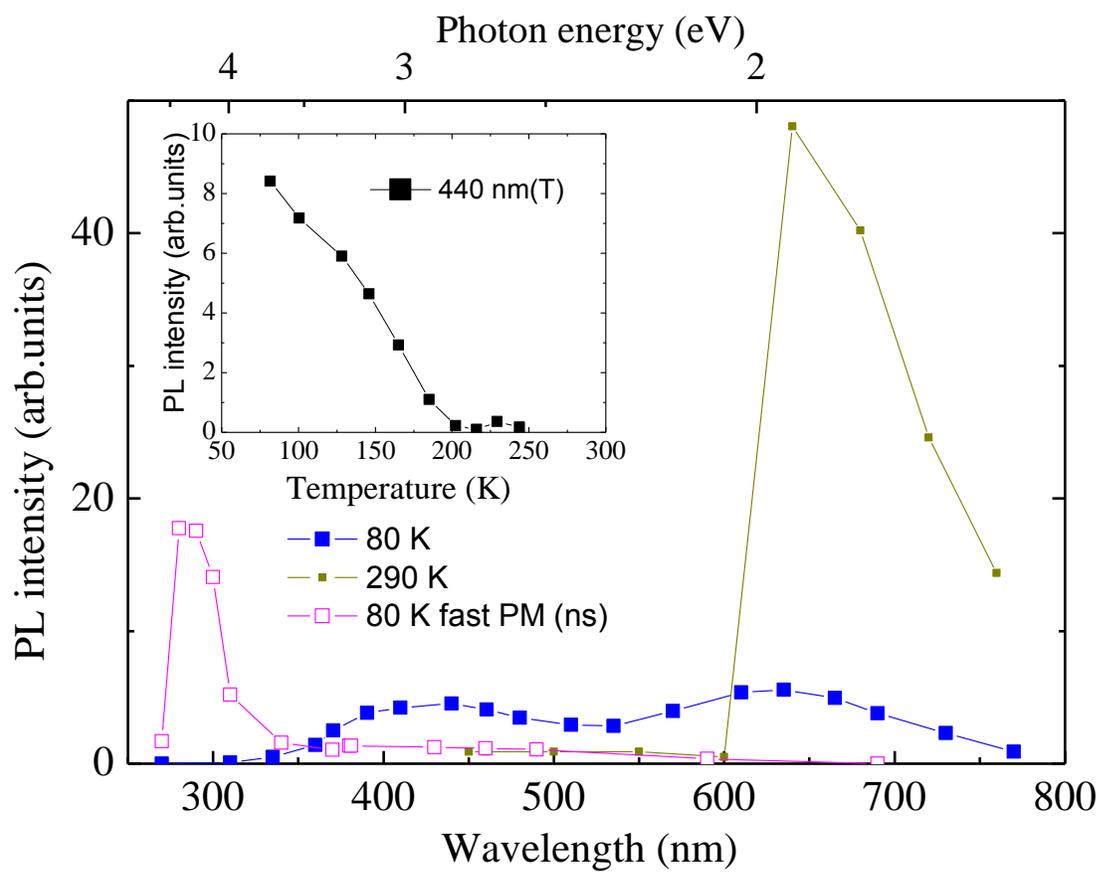

Fig.17

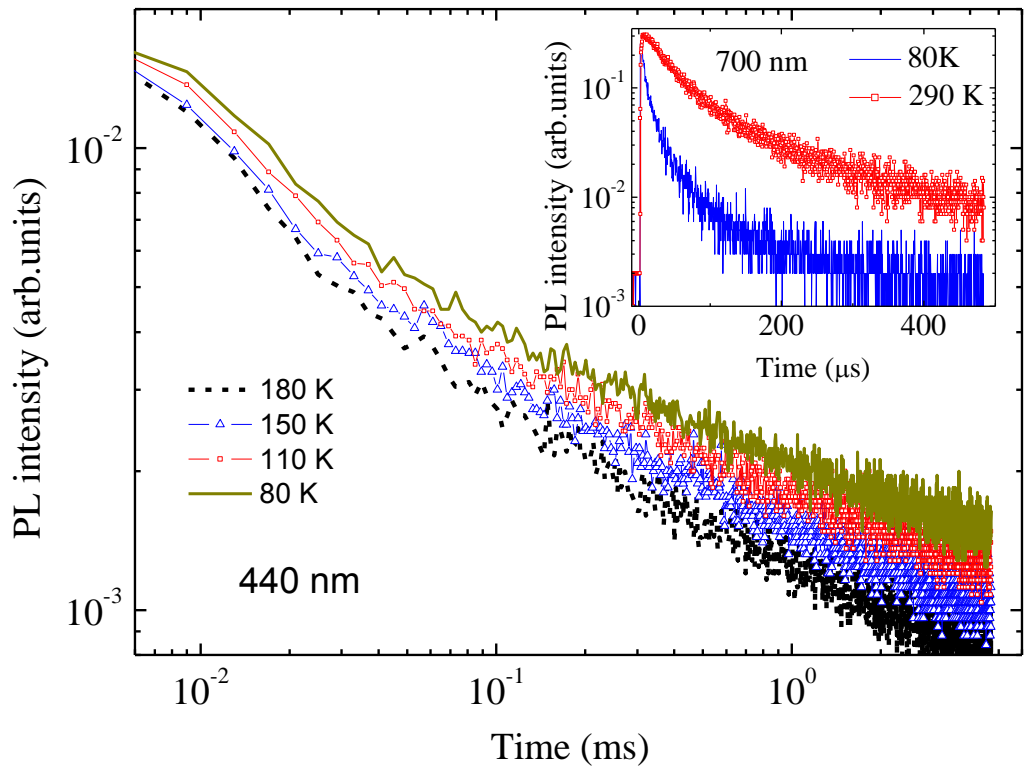

Fig.18